\documentclass[12pt]{article}
\usepackage{epsfig,a4wide}

\newcommand{\be}{\begin{equation}}
\newcommand{\ee}{\end{equation}}
\newcommand{\ba}{\begin{eqnarray}}
\newcommand{\ea}{\end{eqnarray}}

\begin{document}

\title{A critical analysis on deeply bound \\kaonic states in nuclei}

\author{ E. Oset$^{a,b}$ and H. Toki$^b$\\
{\small $^a$Departamento de F\'{\i}sica Te\'orica and IFIC,
Centro Mixto Universidad de Valencia-CSIC,} \\
{\small Institutos de
Investigaci\'on de Paterna, Aptd. 22085, 46071 Valencia, Spain}\\
{\small $^b$ Research Center for Nuclear Physics, Osaka University,
Ibaraki, Osaka 567-0047, Japan}\\
}
\date{\today}
\maketitle

\begin{abstract}
We make a critical analysis on the theoretical calculations that
lead to predictions of deeply bound kaonic states
in nuclei. The model set-up, after dropping several important
processes and channels, leads unavoidably to an unrealistic deep
potential with a very small imaginary part. We
review also the experimental results taken as reference 
for the claim of deeply
bound kaons. We suggest that the peaks of the proton spectra come
from $K^-$ absorption on a pair of nucleons, leaving the rest of
the nucleons as spectators. Based on this conjecture we predict
what would happen in other nuclei.
\end{abstract}

\section{Introduction}
The search for deeply bound hadronic states has a long history.
Deeply bound pionic atoms were predicted theoretically by using various
optical potentials and the widths were expected to be smaller than
the separation between the neighboring levels, which should make a
clear experimental case \cite{toki,toki2,nieves}. The 
 unexpected small widths are due to the presence of a large
repulsive s-wave pion-nucleus interaction, which pushes away pions
from the nuclear interior to avoid pion absorption.  After several
trials, the pionic states were found in the $(d,^3$He) reaction
\cite{expiatom,expiatom2} on the basis of the suggestion given in Ref. \cite{tokidhe,zakidhe} and
far less clearly in the $(\pi^-, \gamma)$ reaction \cite{raywood}
suggested in Ref. \cite{nievesvan}.
The history for the possible $\eta$
bound states is also long
\cite{bhalerao,liu,chiang,sandhas,tsushima,inoue}. Although most optical
potentials lead to bound states, they also share the unpleasant
feature that the widths are wider than the separation of the
levels, which makes the experimental identification difficult. In
spite of this difficult experimental condition, claims have been made that a $\eta$ bound state in
$^3$He can be identified \cite{metag}, although the association of
the experimental signal to a bound $\eta$ state is not completely
unambiguous \cite{meissner}.  Hence, it is important for the
observation of deeply bound hadronic states the existence of a mechanism to
make the widths of the deeply bound hadronic states smaller than
the separation between the neighboring levels.

The other hadron that has attracted attention recently is the
kaon. Kaonic atoms have been phenomenologically studied, and the
large strength of the potential assumed at a time raised hopes
that deeply bound states could exist \cite{galrep}. However, the
imaginary part of the optical potential is large and therefore the
widths of the deeply bound kaonic states come out to be too large
to be detected as distinguished states.  Hence, one should rely on
the microscopic derivation of the kaon optical potential to find a
mechanism for small widths  of deeply bound
kaonic states, if they exist. By deeply bound states we mean states 
bound by about 40-200 MeV. Note that with realistic potentials at small 
densities one obtains in heavy nuclei states bound by a few MeV, which 
have not been observed and which are narrow enough to be distinguishable 
from other levels with the same angular momentum \cite{friedmanlett,friedmannuc,
okumura,baca}.

The microscopic derivation of the
optical potential for kaonic atoms is related strongly to the
properties of the $\Lambda(1405)$ state, which is located just below
the kaon-proton threshold. There have been many studies on the
theoretical derivation of kaon nucleus optical potential. We shall
discuss the details of the recent studies on the $\Lambda(1405)$
state and the density dependence of the kaon-nucleon interaction
in the nuclear medium later on in Sect. 2.

The discussion on the kaons in nuclei became more interesting due to
the prediction made by Akaishi and Yamazaki on the existence of
extremely deep kaonic states \cite{akaishi}. They consider the $\Lambda(1405)$
state as the bound state of the kaon and proton and its width is
caused by the coupling to the pion-$\Sigma(1193)$ channel.
Assuming for the interactions a gaussian form with a fixed width, 
they obtained the interaction parameters by fitting
the mass of the $\Lambda(1405)$ and the kaon-nucleon interaction at 
threshold.  The kaon-nucleon interaction in the isospin I=0 channel comes
out to be strongly attractive and the imaginary part becomes zero
below the pion-$\Sigma$ threshold.  This model set-up leads to
the existence of very deep and quite stable kaonic states 
in $^3$He and $^4$He
systems. At the same time, the large attractive interaction
between the kaon and the nucleons makes the system very compact as much
as ten times the nuclear density. The width becomes then small,
since the deeply bound kaonic states come out to be below the
pion-$\Sigma$ threshold. The width comes solely from a weak
coupling of the deeply bound kaonic states to the pion $\Lambda$ channel
with the isospin I=1.

This proposal of the existence of deeply bound kaonic states,
which have large nuclear density, triggered experimentalists to
detect the kaonic states in light nuclei.  Suzuki et al. made
experiments by using stopped K$^-$ on $^4$He and measured protons
in coincidence with pions \cite{suzuki}.  They identified some
peak structure in the proton spectra. If the peak structure is
identified to be caused by the formation of a strange tribaryon,
the mass is 3115 MeV and the width is about 20 MeV. In recent
talks \cite{sato} it has been  claimed that the strange tribaryon should be
identified with a deeply bound kaonic state although the isospin of
the state is different from the original prediction, since other
possibilities were concluded unlikely to provide the peak
structure in the proton spectrum. The formation probability of
such an exotic strange tribaryon state is surprisingly large, of
order of one percent, for stopped negative kaons, even though the
theoretical model of Ref. \cite{akaishi} requires high density objects for deeply bound
kaonic states, which would lead naturally to small transition nuclear 
matrix elements. 

In this paper, we would like to make a critical analysis of the
theoretical approach, which leads to the prediction of deeply
bound kaonic states in light nuclei and at the same time review
the recent experimental results.  For this purpose, we shall
present the theoretical development of the kaon-nucleon
interaction in free space in the framework of the chiral unitary
model and the modification of the interaction due to the many body effects in 
the nuclear medium in Sect. 2. In the free space (kaon-nucleon system), the 
coupled channel effect of various channels 
leads to a more complex structure for the $\Lambda(1405)$ state.  In the nuclear
 medium, the two body kaon absorption process becomes significant with density.
We shall then discuss the simplified treatment
of the kaon-nucleon interaction and its modification in the
nuclear medium done in Ref. \cite{akaishi} in view of the recent
development of the chiral unitary model in Sect. 3.  In Sect. 4, we
review the experimental results from various different points of view for 
the interpretation of the claimed deeply bound kaonic state.
Sect. 5 will be devoted to the summary.

\section{Kaon-nucleon interaction and the kaon-nucleus
optical potential}

It is very important to understand the kaon-nucleon interaction
and its modification due to the nuclear medium, which leads to the
kaon-nucleus optical potential, for the discussion of deeply bound
kaonic states.  In particular, the understanding of the
$\Lambda(1405)$ is essential in order to study the fate of the
kaons in the nucleus.  The kaon-nucleon threshold is 1432 MeV,
which is 27 MeV above the $\Lambda(1405)$ state.  Hence, the
$\Lambda(1405)$ should get a strong influence from the kaon-proton
interaction.  The width of the $\Lambda(1405)$ is about 50 MeV,
decaying exclusively into the pion-$\Sigma(1193)$ system.  Hence,
the minimum ingredients for the structure of the $\Lambda(1405)$ are
the kaon-nucleon interaction and its coupling to the pion-$\Sigma$
channel. Another feature, which should be taken care for
the development of the kaon-nucleon dynamics, is that the
kaon-nucleon scattering amplitude is repulsive at  threshold.
This feature is well known both from extrapolation of scattering
data \cite{martin}, and from the measurement of kaonic atoms in
the proton \cite{iwasaki}.

By taking a suitable dynamical model, we can make the amplitude
for the $K^- p$ channel to have a typical resonance shape. The peak of
the imaginary part appears about 20 MeV below the $K^- p$
threshold, and the real part changes sign at the same energy,
changing from an attractive interaction below the resonance peak
to repulsion above the peak. As a consequence of it (counting also
the $K^- n$ interaction) the low density $K^-$ optical potential,
$t \rho$,
is repulsive at the $K^- p$ threshold, where $t$ is the kaon-nucleon T-matrix 
and $\rho$ is the nuclear density. The scattering T-matrix
is obtained by solving the Bethe Salpeter equation and hence the
iteration of diagrams shown in Fig. 1 is summed up.

\begin{figure}
\psfig{figure=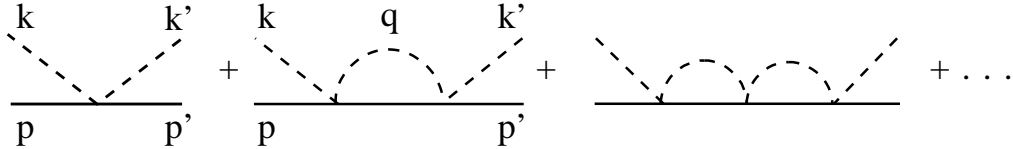} \caption {Diagrammatic representation
of the Bethe Salpeter equations in $\bar{K} N$ scattering, where
the solid line denotes the nucleon and the dashed one the kaon.}
\end{figure}

It is interesting to discuss what happens when the $K^-$ interacts with 
nucleons in a nuclear medium. In the intermediate states (loops
in the scattering series of Fig. 1), there are $\bar{K} N$ states among other 
channels. In the presence of the medium the nucleon intermediate
states have to be placed above the Fermi sea due to the Pauli blocking
effect. This implies that the formation of the $\Lambda(1405)$
resonance requires more energy, as a consequence of which the
resonance shape is reproduced now at higher energies. Hence, as
one can see in Fig.2, when the real part of $t$ is moved at
higher energies, the zero of the resonant scattering amplitude
would cross the $K^- p$ threshold and now the interaction is
attractive at the $K^- p$ threshold. This intuitive picture describes
what happens in the calculations \cite{kochpau} and  is
the main reason to convert the repulsive interaction of the low
density limit into an attractive one as soon as a density
which produces a sensible Pauli blocking builds up.  A potential depth of around
200 MeV for the kaons can be obtained in this way \cite{akaishi}.

We should not stop here, however.  The kaon feels now a strong
attractive potential and the kaon also appears in the intermediate
states of the Bethe Salpeter series. With the strong attraction now on the kaon, its excitation
requires less energy, and hence using the same arguments as
before, the resonance is produced at a smaller energy.  Once again
the zero of the real part of the scattering amplitude moves to
lower energy below the  $K^- p$ threshold, leading once more to repulsion.
 The presence
of the resonance is thus of extreme importance to determine the
fate of the interaction in the medium and, in view of the shift up
and down of the resonance when medium corrections are taken into
account, a reliable calculation requires selfconsistency in the
kaon selfenergy, in the sense that one must determine the kaon
selfenergy, replace it in the kaon propagators, reevaluate the
kaon selfenergy, replace it again in the kaon propagator and so on
till convergence is achieved.

This selfconsistency procedure was done for the first time in Ref.
 \cite{lutz}. It was found that the attraction felt by the kaons
was drastically reduced. Subsequent calculations
\cite{schaffner,galself}, including those which consider also the
renormalization of the intermediate pions in the pion-$\Sigma$
channel \cite{angelsself} were performed and the strength of the
potential was shown to be attractive and of the order of 40-50 MeV
at nuclear matter density. This selfconsistency procedure, including also the
p-wave excitation of $\Lambda h$ and $\Sigma h$ components by the kaon, or 
$ph$ excitation by the pion, automatically 
produces the kaon absorption by two nucleons, $K^- NN \to N \Lambda, N \Sigma$.
  There is hence a reduction of about a factor four for the real part of the 
  kaon optical potential and there appears the imaginary potential due to two 
  body absorption process, when the selfconsistency is implemented.  No
calculation which neglects these important effects caused by the 
selfconsistency procedure should be deemed realistic.

\begin{figure}
\psfig{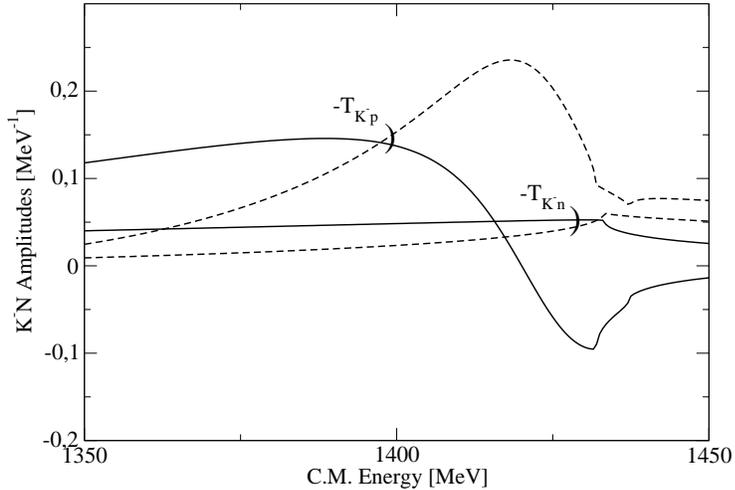}
\caption{Scattering amplitudes for $K^-p \to K^-p$ and $K^-n \to
K^-n$ around and below the $K^-p$ threshold in the chiral unitary
model. The solid curves are the real part and the dashed ones the
imaginary part for K$^-$p and K$^-$n channels \cite{angels}. }
\end{figure}

All the features discussed above are worked out more
systematically in the SU(3)$_f$ chiral unitary model. The
$\Lambda(1405)$ resonance is dynamically generated in the coupled
channel model of $\bar{K} N,  \pi \Sigma$ \cite{dalitz} and in the chiral
unitary model \cite{weisek,angels,ollerulf,manolo,hyodo}, where
one uses input from chiral Lagrangians and, in addition to the
$\bar{K} N, \pi \Sigma$ channels, one also has other channels from
the combination of the octet of pseudoscalar mesons with the octet
of stable baryons.  In these latter works the $\Lambda(1405)$
resonance is generated and a good reproduction of various $K^-p$
reactions is obtained.

The lowest order chiral lagrangian for the meson baryon interaction
is given by
\begin{equation}
L_1^{(B)} = < \bar{B} i \gamma^{\mu} \frac{1}{4 f^2}
[(\Phi \partial_{\mu} \Phi - \partial_{\mu} \Phi \Phi) B
- B (\Phi \partial_{\mu} \Phi - \partial_{\mu} \Phi \Phi)] >
\end{equation}
where $\Phi$ and $B$ are the ordinary SU(3) matrices of the meson and baryon
fields respectively. From there, recalling the dominance of the $\gamma^0$
component at low energies, one deduces the kernel of the Bethe Salpeter
equation (potential V)
\begin{equation}
V_{i j} = - C_{i j} \frac{1}{4 f^2} (k^0 + k'^0)
\end{equation}
where $k^0 , k'^0$ are the energies of the mesons.
The symmetric matrix $C_{i j}$, where $i,j$ stand for the indices of the coupled
channels, $K^- p$, $\bar{K}^0 n, \pi^0 \Lambda, \pi^0 \Sigma^0,
\pi^+ \Sigma^-, \pi^- \Sigma^+, \eta \Lambda, \eta \Sigma^0, \bar{K}^0 \Xi^0,
K^+ \Xi^-$ for charge zero,
is given in Ref. \cite{angels}. The
solution  of the Bethe Salpeter equation, which sums up the diagrams of Fig. 1, is given by
\begin{equation}
T = [1 - V \, G]^{-1}\, V
\end{equation}
in the matrix form of the coupled channels.

\begin{figure}
\psfig{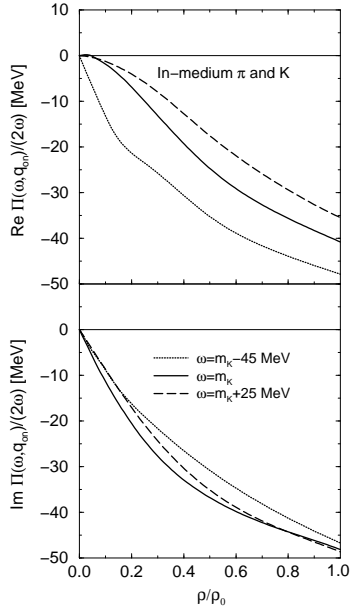} \caption{Real (top) and
imaginary (bottom) parts of the $K^-$ optical potential as a
function of density obtained from the {\it In-medium pions and
kaons} approximation. Results are shown for three different $K^-$
energies: $\omega=m_{K}-45$ MeV (dotted curves), $\omega=m_{K}$
(solid curves) and $\omega=m_{K}+20$ MeV (dashed curves). }
\end{figure}

We note here that one of the surprises of the chiral
approach is that there are not just one $\Lambda(1405)$ but two
states, and the shape obtained in experiments comes from a
superposition of the two resonances with different weights for
different reactions, which makes the position and width of the
resonance vary from one reaction to another \cite{thomas,prakhov}.
The existence of two resonances for I=0, S=-1 was hinted in the
work \cite{ollerulf} and made clear in a more systematic manner in 
Ref. \cite{jido}, where the existence of two octets and one singlet 
of dynamically generated resonances was established. Similar results were
posteriorly found in Ref. \cite{carmina}. The two $\Lambda(1405)$
states would correspond to the mixture of the singlet with one of the
octets, the other octet combination moves up to generate the
$\Lambda(1670)$, which was also reported before \cite{bennhold}.
These two $\Lambda(1405)$ states are quite different, there is one
appearing around 1396 MeV, which has a  width of about 140 MeV and
couples mostly to $\pi \Sigma$, while the other one appears around
1420 MeV, has a width about 30 MeV and couples mostly to
$\bar{K}N$.  These features were also observed in two recent works
which include the effect of higher order chiral Lagrangians
\cite{borasoy,josek}, although the width of the wide resonance is
about 240 MeV in Ref. \cite{borasoy}.

With this elementary input and the corrections to the $\bar{K} N$
amplitude in the medium, the selfconsistent calculation of the
$K^- $ selfenergy in the nuclear medium done in Ref. \cite{angelsself}
gives the results shown in Fig. 3 as a function of the nuclear density. 
The real part of the potential starts from a slightly
positive value (repulsive) at  zero density and switches its sign
to negative (attractive) with increasing density.  The strength of the 
attractive potential becomes larger as the $K^-$ energy is increased.  
The imaginary part is essentially unchanged with the $K^-$ energy.  The 
direct use of this potential for the calculation of kaonic
atoms of light and medium heavy nuclei provides a good
reproduction of the kaonic atom data \cite{okumura}, as shown in Fig. 4.
We can conclude here that the SU(3) chiral unitary model with the
inclusion of all the medium corrections on the baryons and the
mesons is supported from the kaonic atom data. Small diversions from 
the potential of \cite{angelsself} were found in \cite{baca} by performing 
a best fit to the kaonic data. The calculations of \cite{galself} also show 
similar values for the strength of the potential, and this is also the 
case of the $K^-$ optical potential evaluated in \cite{laura}, based on a 
meson exchange model for the elementary $\bar{K}N$ interaction. In this 
latter work, higher partial waves, beyond the dominant s-wave intereraction
were also considered wich had a moderate effect on the potential.
 
\begin{figure}
\psfig{figure=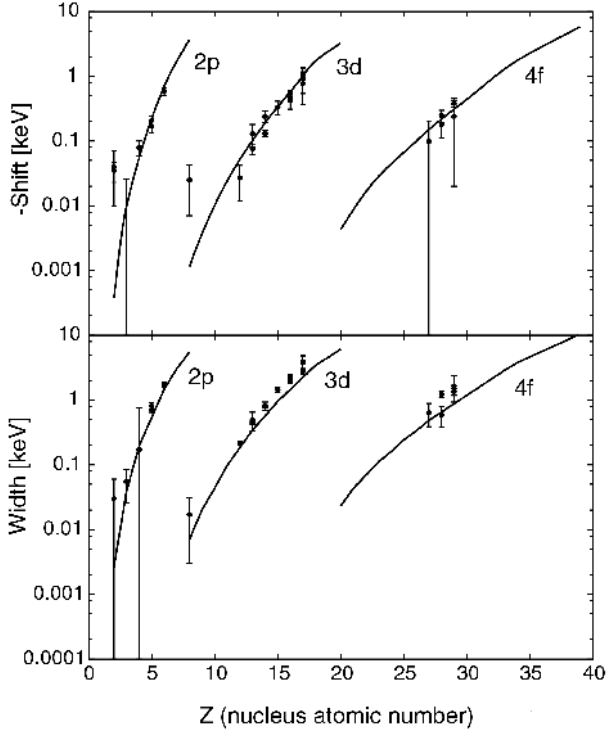,height=10cm} \caption{The energy shifts
and the widths of the kaonic atoms in light and medium heavy nuclei 
\cite{okumura}. The
experimental results are shown by the points with bars, which are
compared with theoretical results obtained by using the optical potential
obtained by the chiral unitary model shown in Fig. 3 \cite{angelsself}. }
\end{figure}

\section{Theoretical claim of deep potential}

In this section we discuss the work of Ref. \cite{akaishi}, where claims
of a deep $\bar{K}N $ potential with a small imaginary part
are made which would provide very stable
deeply bound kaonic states in nuclei.  Quoting the authors
textually, "we construct phenomenologically a quantitative
$\bar{K}N$ interaction model that is as simple as possible using
free $\bar{K}N$ scattering data, the KpX data of kaonic hydrogen
and the binding energy and width of $\Lambda(1405)$, which can be
regarded as an isospin I=0 bound state of $\bar{K}+N$". They use
as input $v_{\bar{K}N,\bar{K}N}$, $v_{\bar{K}N,\pi \Sigma}$,
$v_{\bar{K}N,\pi \Lambda}$, which are fitted to data, and set
$v_{\pi \Sigma,\pi \Sigma}$, $v_{\pi \Lambda,\pi \Lambda}$ equal
zero "to simply reduce the number of parameters".

The assumptions made for the construction of the phenomenological 
$\bar{K}N$ interaction as stated above in Ref. \cite{akaishi} contain 
sufficient elements for a discussion here. As is well known from the 
coupled channel calculations leading to the $\Lambda(1405)$, 
(now we should go back to the two states) 
\cite{weisek,angels,ollerulf,manolo,hyodo}
the appearance of the two $\Lambda(1405)$ states \cite{jido} is a
consequence of an intricate coupled channel dynamics, particularly
the $\bar{K} N$ and $\pi \Sigma$, and none of the states can be
claimed to be a bound state of $\bar{K} N$ as assumed in Ref.
 \cite{akaishi}. In Ref. \cite{akaishi} the rich coupled channel
dynamics is destroyed from the moment that $v_{\pi \Sigma,\pi
\Sigma}$ and $v_{\pi \Lambda,\pi \Lambda}$ are set equal zero.
This is most dramatic when one realizes that the strength of the
$\pi \Sigma \to \pi \Sigma$ in I=0 provided by the chiral
Lagrangians is 4/3 times bigger than that of $\bar{K}N \to
\bar{K}N$ \cite{angels}.  As a consequence, none of the
resulting $\Lambda(1405)$ states qualifies as a bound $\bar{K}N$
state, as one can see by the large coupling of the two states to
both the $\pi \Sigma$ and $\bar{K}N$ channels.  If anything, even
if still unacceptably rough, one should take the narrow state at
1420 MeV, which couples more strongly to $\bar{K}N$, as indicative
of a bound $\bar{K}N$ state, in which case the strength of the
interaction would be much smaller, since the state would be bound
by only 12 MeV, in contrast with Ref. \cite{akaishi} where the binding
energy of the $\bar{K}N$ is assumed to be 29.5 MeV.  As a
consequence of disregarding the coupled channel dynamics and
assuming the $\Lambda(1405)$ as a $\bar{K}N$ bound state with such
a large binding energy, the potential that is obtained from the
fit is very large to begin with, the strength of the potentials at
$r=0$ is of the order of 400 MeV and the range of the interaction
is also very small, of the order of 0.66 fm, which forces
artificially the kaons to stick to the nucleons, producing nuclear
densities with three baryons of the order of ten times the nuclear
matter density at the center of the three baryon system in the
calculation of Ref. \cite{akaishi}.

This discussion indicates that the approximations done in 
Ref. \cite{akaishi} necessarily lead to larger scattering amplitudes 
than the chiral approaches. This is indeed the case as one can see by
comparing Fig. 9 of Ref. \cite{angels} with Fig. 1 of Ref. \cite{akaishi}.
To facilitate the comparison, the I=0 amplitude plotted in Fig. 1
of Ref. \cite{akaishi} is equivalent to

\begin{equation}
f^{I=0}= - \frac{1}{4 \pi} \frac{M}{\sqrt{s}} (2 t_{K^- p}- t_{K^-n})
\label{i0amp}
\end{equation}
with $t$ the amplitude used in Ref. \cite{angels}.

We can see that in the plateau around 60 MeV below the $\bar{K}N$
threshold the chiral amplitude is about 2.4 fm in Ref. \cite{angels}, 
while that of Ref. \cite{akaishi} is about 5 fm. One may argue that the 
 $\bar{K}N$ amplitude below the threshold is not an observable and 
 one would like to have a closer experimental evidence of the failure 
 of the model of Ref. \cite{akaishi}.  Such an observable can be seen in 
 the recent
experiment of the reaction $K^- p \to \pi^0 \pi^0 \Sigma$ in the
region of excitation of the $\Lambda(1405)$ \cite{prakhov}.  As
one can see in the experiment the peak of the $\Lambda(1405)$
appears at 1420 MeV, while the one of Ref. \cite{akaishi} appears at
1403 MeV, the nominal value of the $\Lambda(1405)$ mass.  The
explanation given in Ref. \cite{magas} of the surprisingly large
$\Lambda(1405)$ mass
seen in Ref. \cite{prakhov} was natural within the context of the two
$\Lambda(1405)$ states obtained in the chiral theories. Indeed,
since the high energy $\Lambda(1405)$ state around 1420 MeV
couples mostly to $\bar{K}N$ the reaction $K^- p \to \pi^0 \pi^0
\Sigma$ gives more weight to this state in the amplitude,
reflecting the most efficient mechanism to create the
$\Lambda(1405)$ which is the emission of a $\pi^0$ prior to the
scattering $K^- p \to \pi \Sigma$. Obviously, a theory with a
unique $\Lambda(1405)$ at 1403 MeV would not reproduce
 this experiment.

The $\bar{K}N$ interaction of Ref. \cite{akaishi} explained before is used to 
calculate the $g$ matrix in the nuclear medium,
which is then used as a $K^-$ nucleus optical potential.  The Pauli
blocking on intermediate states is implemented there and this
converts the initial interaction into an attraction as we have
explained above, only the strength obtained for kaon selfenergies
around threshold is about 3.5 times bigger than the one obtained
in Ref. \cite{angelsself,galself,schaffner}. The starting energy in the
$g$ matrix calculation is changed to smaller values to account for
the possibility of having kaons with smaller energy than free
kaons (this is also done in Ref. \cite{angelsself}, where the $K^-$
selfenergy is calculated for several values of the starting $K^-$
energy as shown in Fig. 3.). However, this binding energy is not used in the 
intermediate states in the scattering equations 
in Ref. \cite{akaishi}
and the important effects of the selfconsistency explained above
are simply lost, leading to an unrealistically large attractive
potential.

In Ref. \cite{akaishi} this deep potential is used to explore the
possibility of finding $K^-$ bound states in light nuclei with A=3, and 4. 
The same equation as in nuclear matter is used with an average
Fermi momentum. It is easy to see from the equations used in section 3A of 
Ref. \cite{akaishi} that the Fermi momentum used there is equal or larger than 
that of nuclear matter at normal nuclear matter density.  It is not intuitive 
that three or four nucleon systems have the same Pauli blocking effect than a 
full nuclear matter environment, but with this approximation and the former 
ones that lead to the unrealistically deep potential as we discussed, the 
potential for the A=3 and 4 systems is
evaluated which leads to binding energies of the kaon of the order
of 70 MeV with widths around 75 MeV, the imaginary part coming
from $K^- N \to \pi \Sigma$.

At this point the authors argue that since the attraction due to
the kaon is so large, it can induce a further contraction of the
nucleus, increasing its density and consequently leading to more
attraction on the kaon till the nuclear incompressibility stops
the system from shrinking even further. With this mechanism,
densities of the order of ten times the nuclear matter density in
the center of the nucleus are obtained in Ref. \cite{akaishi}.  This
finding should be strongly questioned since to arrive to it
they use the effective NN interaction \cite{hasegawa}, which
accounts for the incompressibility of nuclear matter.  Yet, at these
high densities, and even much before, nuclear matter does not
resemble this simple picture, and without entering into more
complicated scenarios it is enough to quote that excitation of
strange baryons would become favorable when matter is compressed
with only three times or so of the nuclear matter density
\cite{dover,bombaci,vidana}.  In this direction it is also worth pointing out
that, as matter compresses, the effect of three body forces becomes more
relevant introducing a repulsion that leads to a stiffer equation of state
 ( a recent evaluation of such three body forces, that essentially produce a
 repulsive effect, can be seen in \cite{murat}). It has been pointed out that 
 with a stiffer equation of state the $K^-$ optical potential gets weekened
 \cite{ maruyama,greiner}.

With these calculations the authors of Ref. \cite{akaishi} obtain deeply bound 
kaonic states bound by 108 MeV in $^3$H and 86 MeV in $^4$H, with
the widths of respectively 20 and 24 MeV due to the $ \pi \Lambda$ decay
channel since now the strong $\pi \Sigma$ decay channel is closed.
This last issue opens a new discussion on the width of these states. 
In Ref. \cite{akaishi} the widths are considerably
reduced by forcing the states to appear below the $\pi \Sigma$
threshold. Yet, in nuclear matter the two decay channels mentioned
above are not the only ones and two body decay channels appear,
like $K^- NN \to \Lambda N,  \Sigma N, \Sigma(1385) N$ . These two
body decay channels and other multinucleon decay channels are
automatically taken into account in the selfconsistent calculation
of Ref. \cite{angelsself}, but not in Ref. \cite{akaishi}. Instead of explicitly 
calculating the two body absorption processes, they are estimated in 
Ref. \cite{akaishi} by taking the empirical value of 16.5
percent branching ratio for the non pionic $K^-$ absorption in
$^4$He \cite{katz}. Then the authors apply the fraction of 17
percent to the imaginary part of the potential calculated in
Ref. \cite{galself} which produces a contribution of 11 MeV in the
imaginary part of the potential for nuclear matter density,
hence leading to a width of 22 MeV. However, the authors quote an
approximate value of 12 MeV for the two small nuclei. Here comes
an observation to make which is that these two body mechanisms
calculated in Ref. \cite{angelsself} are proportional to the nuclear
density squared and hence, strictly speaking, at densities ten times
the nuclear matter density as claimed in Ref. \cite{akaishi} would
become 100 times larger, which means, inducing widths of 2000 MeV.
Obviously this is a gross overestimate but invoking the "average"
density of about three times nuclear matter density of
Ref. \cite{akaishi} one should multiply this kaon absorption mechanism
by a factor around ten, thus rendering the width of about 200 MeV, which is 
larger than the binding energy.

In the discussion mentioned above we have exploited the inconsistencies 
resulting from the chain of rough approximations done in Ref. \cite{akaishi} 
leads to.  By contrast to that, the
realistic calculations done in Ref. \cite{schaffner,angelsself} lead to deeply 
bound states in nuclei of the order of 10 and 30 MeV in $^{40}$Ca \cite{okumura},
 but with the width of the order of 100 MeV.  An independent work in 
 Ref.\cite{galabs} finds that in the event that bound states of kaons existed in 
 medium nuclei in the region
of 100-200 MeV, the widths would not be smaller than about 40 MeV and these 
widths would be even larger in light systems with the claims of larger 
densities made in Ref. \cite{akaishi}.

The developments of Ref. \cite{akaishi} have a continuation after the 
performance of an experiment at 
KEK \cite{suzuki}. This experiment claims to have found a strange tribaryon 
S(3115) which is seen as a peak in the spectra of emitted protons following 
the absorption of stopped $K^-$ in $^4$He. The authors of Ref. \cite{suzuki} do 
not mention in the paper the possibility that this state is the kaon bound 
system obtained in Ref. \cite{akaishi}, since it would correspond to a $K^-$ 
bound by 195 MeV and it has necessarily I=1, while the one predicted in 
Ref. \cite{akaishi} has a binding energy of 108 MeV and is I=0. Yet, in view of 
this experimental finding, the authors of Ref. \cite{akaishi} modify their 
approach by introducing relativistic effects, some contribution from spin-orbit
 interaction plus an {\it ad hoc} increase of the bare $\bar{K} N$ interaction 
 by 15 percent, till a new $K^-$ bound state by 195 MeV appears \cite{akainew}.  
 Let us quote, in connection to this, that in the chiral calculations of the 
 optical potential of kaons in the medium of Ref. \cite{lutz,angelsself,schaffner}
the kaons were always treated relativistically and the study of the deeply 
bound $K^-$ states in nuclei was done also relativistically using the Klein 
Gordon equation \cite{okumura}, as now used in Ref. \cite{akainew}.   It is also 
worth quoting that
after all the original rough approximations that induce further unreliable 
corrections from the nuclear shrinkage and the new relativistic corrections, 
the optical potential has acquired a strength of 618 MeV, with an imaginary 
part of 11 MeV. In contrast, the chiral calculations give a strength of 40-50 
MeV, which would not give room for a further shrinkage of the nucleus since 
this is the strength of the nucleon nucleus potential, and the addition of a one
more nucleon to a nucleus barely modifies its density. Furthermore, they have an 
imaginary part sizably larger, of the order of 50 MeV, due to the unavoidable 
many body decay channels.

\section{Discussion on the "experimental strange tribaryon state"}

On the basis of the discussion in the former section it is quite clear that 
the KEK experiment \cite{suzuki} could not be
interpreted in terms of the creation of deeply bound kaonic states. From
this perspective we would like to interpret the meaning of the peak seen there.
The experiment is 
\begin{equation}
stopped~ K^- +~^4He \to S +~p
\label{triprod}
\end{equation}
where S has a mass of about 3115 MeV, and has the quantum numbers
of $YNN$ with zero charge where $Y$ is a S=-1 hyperon. The state S
would have $I_3=-1$ and hence cannot be $I=0$ as predicted
originally in Ref. \cite{akaishi}, which was already noted in
Ref. \cite{suzuki}. The peak is also quite narrow, around or smaller
than 20 MeV. A tempting idea could be a bound state of $\pi
\Lambda NN$, which, as noticed in Ref. \cite{suzuki}, is only about 16
MeV above the observed peak. Of course, after what has been
learned in chiral unitary models \cite{ourreport}, one would have
to study this with coupled channels, which would make the
identification as such a state possible only at a qualitative
level. The width would come from pion absorption on two nucleons, which is expected large \cite{ashery,futami}.

There are, however, other potential explanations of the peak, which
would deserve some attention.  One of them was already pointed out
in Ref. \cite{suzuki}. The peak of the proton comes at about 500 MeV/c.
There is a process that can create a peak at the same place. This
is the reaction
\begin{equation}
K^- + ^4He \to ^4_{\Lambda}He + \pi ^-
\label{twostep1}
\end{equation}
followed by
\begin{equation}
 ^4_{\Lambda}He  \to ^3H +p
 \label{twostep2}
\end{equation}
In this chain a pion  with 255 MeV/c and a proton of 508 MeV/c
would be emitted.  This possibility is discarded in Ref. \cite{suzuki}
in view of the following arguments:  

1) The yield of this reaction
is estimated to be of the order of $2 \times 10^{-4}$ according to Ref.
 \cite{outa}, while the rate for the reaction of Eq.~(\ref{triprod}) is also 
 estimated in Ref.\cite{suzuki} to be of the order of one percent.
Actually, we do not find in Ref. \cite{outa} the estimate claimed in
Ref. \cite{suzuki}, and in Ref. \cite{suzuki} no reasons are given either
for the yield of one percent claimed for the reaction of
 Eqs.~(\ref{twostep1},\ref{twostep2}),
 although, based on the experimental yields for hypernuclear
formation \cite{tamura} and the weight of the proton peak in
Ref. \cite{suzuki} these assumptions look reasonable.  

2) The authors
of Ref. \cite{suzuki} make also a  test to eliminate the
possibility of the chain reaction by looking at the secondary
pions.  There a "fast $\pi$" cut is made which would accept 90
percent of the 255 MeV/c pions. If the proton peak came from the
chain reaction it would appear in connection with the "fast $\pi$"
cut. However, the peak is seen both in the "fast $\pi$" cut case
as well as in the complementary set of data with about the same
intensity, which would exclude the interpretation of the peak as
coming from the chain reaction.  There could be a caveat in the
former argument since the experiment is using a thick target of
superfluid helium of 15 cm long and 23.5 cm of diameter at a
density of 0.145 g/cm$^3$. Indeed, the pions of 255 MeV/c are on
top of the $\Delta$ resonance region and hence the cross sections
are large. In fact, by looking at the experimental cross sections
of the paper of Ref. \cite{ashery} one induces a quasielastic cross
section for $\pi^- ~^4$He in the region of interest of about 190
mb. It is easy then to estimate that about five percent of the pions
would undergo quasielastic collisions and due to Pauli blocking
they would lose energy because they need to transfer momentum to
the nucleons to excite them on top of the Fermi sea. The result of
this exercise, however, is that not enough pions would lose energy
to make 50 percent of the signal appear outside the "fast $\pi$"
cut. Hence, based on the tests made in Ref. \cite{suzuki} and the 
former discussion, we do
accept their conclusion that the peaks are not due to the chain
reaction of Eqs.~(\ref{twostep1},\ref{twostep2}). Nevertheless, it
would be very interesting to have the pion spectrum in coincidence with the 
proton peak in order to learn more about the reaction.

The former reflexions gain a new dimension when one realizes that
a similar proton peak is seen in the experiment at FINUDA in
different nuclei from $^6$Li at around 510 MeV/c
\cite{piano}.  While the interpretation of these states as deeply
bound kaonic states would lead to the unlikely result that the
$K^-$ binding energy is about the same for different nuclei from
three to five to more baryons, the alternative explanation as creating
strange multibaryon system with the same binding energy for all
nuclei, would not be less surprising. Such systematics in
different nuclei deserve a serious thought to see a clear
explanation. Certainly, a chain reaction with the formation of a
$\Lambda$ hypernucleus and a posterior decay through nonmesonic
decay into a proton and a bound normal nucleus could, a priori, be
a likely explanation for the peaks, but the tests done in
Ref. \cite{suzuki} give not much room for hope in this direction.
Nevertheless, it would still be interesting to measure the
spectrum of the pions in coincidence, for the reasons given above.

\section{Suggested mechanism}
In view of the previous unsuccessful trials let us explore the
reaction which is most likely to happen: This is  $K^-$
absorption by two nucleons in $^4$He leaving the other two
nucleons as spectators. This kaon absorption process should happen
from some $K^-$ atomic orbits which overlap with
the tail of the nuclear density and hence the Fermi motion of the
nucleons is small.  Then we would have $K^- NN \to \Lambda N$,
and $\Sigma N$, and the two baryons are emitted back to back with
the momentum for the proton of 562, and 488 MeV/c respectively.
These results are very interesting: the peak of the proton
momentum in Ref. \cite{suzuki}, before  proton energy loss corrections,
appears at 475 MeV/c (see Fig. 5 of this reference). This matches
well with the 488 MeV/c proton momentum from a  $K^- NN \to \Sigma
N$ event, and the proton would  lose about 13 MeV/c when crossing
the thick target. This energy loss is compatible with the
estimate of about 30 MeV/c in Ref. \cite{suzuki}, particularly taking
the width of the peak also into account.

This suggestion sounds good, but then one could ask oneself:  what
about $K^- NN \to \Lambda N$? Should not there be another peak
around 550 MeV/c, counting also the energy loss? The logical
answer is yes, and curiously one sees a second peak around 545
MeV/c in the experiment. The peak is clearly visible although less
pronounced than the one at 475 MeV/c and it appears in the region
of fast decline of the cross section.

There are other arguments supporting our suggestion. Indeed, as
mentioned above, the pion momenta from $\Lambda$ decay are smaller 
than those from $\Sigma$ decay. As a consequence of this
we should expect the peak associated with $p \Lambda$ emission to
appear in the low momentum side of the pion (the range of pion momenta is from
61 MeV/c to 146 MeV/c from phase space considerations). Actually, this is the
case in the experiment of Ref. \cite{suzuki} as one can see in Fig. 5d
of this reference, corresponding to the spectrum when the low pion cut is
applied, where the peak of higher momentum stands out more clearly.
 On the other hand, by working out the phase space for
$\Sigma$ decay, the pion momenta range from 162 Mev/c to 217 MeV/c and the pion 
could be
seen in the two regions of pion momenta  of Ref. \cite{suzuki}, as it is indeed 
the case (see figs. 5c, 5d of Ref. \cite{suzuki}). 

One can even argue about the size of the peaks and their relative
strength. For this the information of
Ref. \cite{katz} is very useful. There we find the following results
for events per stopped $K^-$:

\begin{eqnarray}
\Sigma^- p~d ~~~~~1.6 ~\%
\label{sigma1}\\
\Sigma^- ppn  ~~~~2.0 ~\%
\label{sigma2}\\
\Lambda (\Sigma^0) pnn ~~~~11.7 ~\%
\label{sigma3}
\end{eqnarray}
with errors of 30-40 \%.

For the separation of the $\Lambda (\Sigma^0)$ contribution a
rough approximation is done in Ref. \cite{katz}, assuming that the
$\Sigma^0$ yield is one half of the sum of that of  the charged
$\Sigma$ invoking isospin symmetry. In this case the 11.7 percent
of Eq.~(\ref{sigma3}) is split between 2.3 percent  for $\Sigma^0$ 
and 9.4 percent for
$\Lambda$. Isospin symmetry is not that simple with the system of
four baryons in the final state and we estimate uncertainties by
applying a different method. We can assume $K^- pp \to \Sigma^0
(\Lambda) p$ proceeding via $K^- p \to \Sigma^0 (\Lambda) \pi^0$
with the pion virtually exciting particle-hole states of the proton type.
In this case we would have approximately the yields Y of the two
processes as,
\begin{equation}
\frac{Y~(\Sigma^0 NN)}{Y~(\Lambda NN)}=\frac{\sigma(K^- p \to
\pi^0 \Sigma^0)} {\sigma(K^- p \to \pi^0 \Lambda)}
\label{ratio}
\end{equation}

Next we use experimental information from Ref. \cite{kim} to estimate
the ratio of Eq.~(\ref{ratio}). For a $K^-$  of 100 MeV/c momentum 
the two
cross sections are of the order of 20 mb. Using this information
we count how many times we have $\Sigma p$ and $\Lambda p$ in
the final state after $K^-$ absorption ( in the case of $\Sigma^-
p~d$ we naturally associate the $d$ with the spectator block) to
find the probability to have a $Yp$ being the emitted pair, and we
get about 3.7 percent for $\Sigma p$ and 3.1 percent for $\Lambda
p$ when we use the partition of $\Sigma^0 (\Lambda)$ from
\cite{katz} and 4.88 percent versus 1.95 percent when we use Eq.~(\ref{ratio}),
 with experimental errors of the order of 50 percent or more.
The absolute values for $\Sigma p$ emission make sense from the
perspective that in Ref. \cite{suzuki} a rate of one percent for the
production of the peak is estimated. A rate of one in four for the
back to back emission of the pair and spectator remnant pair is
realistic. On the other hand we see, that even with admitted large
uncertainties, ratios of a factor 1.2 to 2.5 for $\Sigma p$ versus
$\Lambda p$ come up from analysis of present data. A larger
strength for the peak of lower proton momentum( $\Sigma p$
emission) relative to that of the smaller proton momentum
($\Lambda p$ emission), of about a factor of two, is actually seen
in Fig. 5(d) of Ref. \cite{suzuki}. The $\Sigma^- p~d $ final state of
Eq.~(\ref{sigma1}), with  the spectator $d$ forming a bound state, should be
one of the largest contributors to the $\Sigma^- p$ peak, leading
to a momentum of the proton of 482 MeV/c.

The hypothesis advanced should have other consequences.  Indeed,
this peak should not be exclusive of the small nuclei. This should
happen for other nuclei. Actually in other nuclei, let us say $K^-
~ ^7$Li, the signals that we are searching for should appear when
a proton  as well as a $\Sigma$ or a $\Lambda$ would be emitted
back to back and a residual nuclear system remains as a spectator and
stays nearly in its ground state.
We would thus expect
two new features: first, the two peaks should be there. However,
since now the spectator nuclear systems remain nearly in their
ground states, only about the
binding energy of the two participant nucleons will have to be
taken from the kaon mass, instead of the 28 MeV in $^4$He for a
full break up, as a consequence of which the proton momenta should
be a little bigger. We make easy estimates of 502 MeV/c for the
proton momentum in the case of $p \Sigma$ emission and 574 MeV/c
for the case of $p \Lambda$ emission.  Curiously the FINUDA data
\cite{piano} exhibits two peaks in $^7 Li$ around 505 MeV/c and
570 MeV/c.

There is one more prediction we can make. The process discussed
has to leave the remnant nucleus in nearly its ground state.
This means
that one has to ensure that the nucleus is not broken, or excited
largely, when the energetic protons go out of the nuclear system.
Theoretically one devises this in
terms of a distortion factor that removes events when some
collision  of the particles with the nucleons takes place.
Obviously this distortion factor would reduce the cross sections
more for heavier nuclei and, hence, we should expect the signals
to fade away gradually as the nuclear mass number increases. This
is indeed a feature of the FINUDA data \cite{piano}.

 Similarly, we can also argue that the spectator nucleus, with a momentum equal 
 to that of the combined pair on which $K^-$ absorption occurs, will have smaller
 energies for heavier nuclei since their mass is larger. Hence, the spreading of
 the energy of the emitted proton should become narrower for heavier nuclei. 
 This is indeed a feature of the FINUDA data \cite{piano}.

This sequence of predictions of our  hypothesis,
confirmed by the data of \cite{suzuki} and \cite{piano}, provides
a strong support for the mechanism suggested of $K^-$ absorption
by a pair of nucleons leaving the rest of the nucleons as
spectators. Certainly, further tests to  support this idea,
or eventually refute it, should be  welcome. An
obvious test is to search for $\Sigma$ or $\Lambda$ in
coincidence and correlated back to back with the protons of the peak.
 Awaiting further experimental
information, we can say that after showing that the
theoretical basis for the kaonic atoms hypothesis was a
consequence of rough approximations which produced potentials one
order of magnitude bigger than any realistic calculation, and
after using information of the experiment of Ref. \cite{suzuki} which
ruled out a likely two step process, by elimination we reached the
present conclusion that, as we could prove, passes all tests of
the experimental information provided by the KEK and FINUDA data.

\section{Conclusions}
In this paper we have made a thorough review of the theoretical
developments that led to predictions of deeply bound kaonic atoms
in light nuclei.  We could show that there were many
approximations done, which produced unreliably deep potentials.  Two main reasons made the
approximations  fail dramatically:  the problem of the coupled
channels to produce the two $\Lambda(1405)$ states was reduced to
only one channel, the $\bar{K}N$, and only one $\Lambda(1405)$,
 which was assumed to be
a bound state of the $\bar{K}N$ potential was considered.   
The second serious problem was the lack of
selfconsistency in the intermediate states, which makes the
results absolutely unreliable when one is in the vicinity of a
resonance, as is the case here. There were many other 
approximations but the former two are sufficient to obtain a
potential as large as ten times what one gets without making these
approximations.  We also state  that the width becomes zero in the
I=0 channel due to the binding energy being lower than the
pion-$\Sigma$ threshold.  However, the selfconsistency
consideration automatically produces the two body kaon absorption
processes, which provides large widths and increases as the density
squared.

With the weakness of the theoretical basis exposed and the realization that
binding energies of 200 MeV for a kaon in a system of three
particles are out of scale, we looked for a plausible
explanation of present experiments which could be interpreted
differently than creating these deeply bound $K^-$ states.
After using information and the analysis of Ref. \cite{suzuki} discarding
potential alternatives with chain reactions, we were led by
elimination to a  source of explanation deceivingly simple, and
which, however,  passes the present experimental tests: the
association of the observed proton peaks to $K^-$ absorption by a
nucleon pair leaving the rest of the nucleons as spectators. From
the emission of $p \Sigma$ we explained the peak found in
Ref. \cite{suzuki} and from the emission of $p \Lambda$ we predicted
another peak which is indeed present in the experiment of
Ref. \cite{suzuki}.  The hypothesis made led us to conclude that these
peaks should also be visible in other nuclei at slightly larger
proton momenta, should be narrower and their strength should
diminish with increasing mass number of the nucleus. Fortunately,
all these predictions could be tested with present data from
FINUDA \cite{piano} and all the predictions were confirmed by
these preliminary data.

Although further experiments could be advisable to further test
our claims, it is also clear that the evidence provided here for
these claims is by far larger than the one for deeply bound kaonic
atoms, based exclusively on theoretical predictions that we have
proved here to be largely overblown. Not to mention that the real
predictions gave a binding energy half the one of the peak of 
Ref. \cite{suzuki}, which would be assumed as a bound state of a kaon, 
and with the wrong isospin. The {\it a posteriori} corrections
of the theory to match the experimental findings and increase the
binding energy by a factor of two only added more uncertainties to
the already unacceptably rough approach on which the genuine
predictions were made.

\section*{Acknowledgments}
We are grateful to Tadafumi Kishimoto and Takashi Nakano for fruitful 
discussions.  We appreciate a 
careful reading of the present manuscript by Mamoru Fujiwara, Manuel Jose
Vicente Vacas and Angels Ramos.  This work is 
partly supported by DGICYT contract number BFM2003-00856,
and the E.U. EURIDICE network contract no. HPRN-CT-2002-00311.
This research is part of the EU Integrated Infrastructure Initiative
Hadron Physics Project under contract number RII3-CT-2004-506078.


\end{document}